# Fermi level equilibration at the metal–molecule interface in plasmonic systems

*Andrei Stefancu[1], Seunghoon Lee[2], Li Zhu[3], Min Liu[3], Raluca Ciceo Lucacel[1], Emiliano Cortés[2\*], Nicolae Leopold[1\*]*

[1] Faculty of Physics, Babeș-Bolyai University, 400084 Cluj-Napoca, Romania

[2] Chair in Hybrid Nanosystems, Nanoinstitute Munich, Faculty of Physics, Ludwig-Maximilians-Universität München, 80539 Munich, Germany

[3] State Key Laboratory of Powder Metallurgy, School of Physics and Electronics, Central South University, 410083 Changsha, China

Corresponding e-mail address:

(E.C.) Emiliano.Cortes@lmu.de;

(N.L.) Nicolae.Leopold@ubbcluj.ro;

**Abstract:** We highlight a new metal-molecule charge transfer process by tuning the Fermi energy of plasmonic silver nanoparticles (AgNPs) *in-situ*. The strong adsorption of halide ions upshifts the Fermi level of AgNPs by up to ~0.3 eV in the order: $Cl^-<Br^-<I^-$, favoring the spontaneous charge transfer to aligned molecular acceptor orbitals until charge neutrality across the interface is achieved. By carefully quantifying experimentally and theoretically the Fermi level upshift, we show for the first time that this effect is comparable in energy to different plasmonic effects such as the plasmoelectric effect or hot-carriers production. Moreover, by monitoring *in-situ* the adsorption dynamic of halide ions in different AgNP–molecule systems, we show for the first time that the catalytic role of halide ions in plasmonic nanostructures depends on the surface affinity of halide ions compared to that of the target molecule.

**Introduction**

Plasmonic photocatalysis has recently emerged as a new paradigm in the sunlight-to-chemical energy conversion cycle.[1-7] However, our physicochemical understanding of the metal-molecule interface and its reactivity is still in its early stages.[8-11] We can gain further insight on how to manipulate and tune metal interfaces at the nanoscale by borrowing some concepts from more advanced fields such as heterogeneous catalysis, photocatalysis, electrocatalysis and surface science.[12-14] Let us focus on one of the simplest examples: a charge transfer process across the metal–molecule interface.

A molecule's donor and acceptor states are its occupied (HOMO) and unoccupied (LUMO) frontier orbitals, respectively. During chemisorption, the frontier orbital energies change relative to the free state (unbound molecule). The energy change of a molecule's frontier orbital when adsorbed to a surface, named energy level alignment, can either enable or inhibit *spontaneous* charge transfer with the solid.[15-18] Thus, during the chemisorption of a molecule on a metal substrate, charge transfer can occur between the metal and the molecule, if the donor and acceptor energy states of the two partners are aligned. This process ends up changing the charge state of the adsorbed molecule (i.e., promoting its reduction or oxidation)[15, 17-18]. Accordingly, the surface metal atoms can be reduced or oxidized to equilibrate the energy levels in the metal–molecule system (i.e., redox potentials). This has been recently shown in dissolution experiments with silver nanoparticles (AgNPs).[19-21]

Even though the effect of tuning the Fermi energy of metal nanostructures through ligands has been known since the early 90s through the work of Henglein and coworkers,[22-24] it has not been explored extensively in the fields of plasmonic chemistry or surface-enhanced Raman scattering (SERS), being more commonly seen in the fields of molecular electronics[25] or quantum dots.[16, 26-27] This concept could have a strong impact in plasmonic catalysis, as both the Fermi level of the NPs and the energy of the absorbed photon determine the energy of the generated hot-carriers.[28-31]

Here, we show a simple and straightforward approach to tune the Fermi level of AgNPs through the adsorption of $Cl^-$, $Br^-$ or $I^-$, which can block or promote the charge transfer to an adsorbed molecule. As a test system, we used the well-known adsorption of methylene blue ($MB^+$) on AgNPs.[32-36] We show that it is possible to reduce $MB^+$ to the $HMB^{+\bullet}$ form only by controlling and tuning the surface chemistry. Furthermore, we used SERS to monitor the surface dynamics of our target analyte, $MB^+$, and of halide ions, which allowed us to gain a mechanistic understanding on the different adsorption regimes of halide ions on the metal surface. The novelty of our study consists in carefully quantifying the surface potential shift of AgNPs upon the addition of halide ions both experimentally (through XPS and UV-Vis spectroscopy) and theoretically (through DFT calculations). Therefore, we show here that the shift of surface potential upon the addition of halide ions is comparable to other plasmonic effects, such as light-induced voltages (photopotential), for which surface potential shifts of $\sim 0.1$ eV were reported,[37-38] or the energy of plasmon derived hot-carriers participating in certain chemical reactions.[13, 31, 39] We believe that the method shown in the present study could open-up new avenues not only for chemical reactions at the nanoscale, but also for connected areas, such as plasmoelectronics.

**Results and Discussions**

**$MB^+$ reduction.** To track the energy level equilibration at the metal–molecule interface, we chose the reduction of $MB^+$ as a model system. $MB^+$ has a two steps electrochemical reduction pathway (Figure 1A); the first one, to the intermediate form, $HMB^{+\bullet}$, takes place on Ag surfaces at a relatively low potential: –0.27 V (vs. Ag/AgCl); while the heterogeneous reduction to leuco methylene blue form (LMB) takes place at –0.365 V (vs. Ag/AgCl).[36, 40]

Through the reduction of $MB^+$ to the intermediate radical $HMB^{+\bullet}$ and LMB forms, respectively, the molecular optical resonance shifts further away from the visible to the UV range (Supporting Information, Figure S1).[40-41]

To probe the SERS features of each $MB^+$ form (i.e. $MB^+$, $HMB^{+\bullet}$ and LMB), we used a highly potent reducing agent, $NaBH_4$, which was shown to fully reduce $MB^+$ (Supporting Information, Figure S2).[42] The newly formed N-H bond in both $MB^+$ reduction products, $HMB^{+\bullet}$ and LMB, is observed at ~1130 $cm^{-1}$.[40] Additionally, the reduction of $MB^+$ to $HMB^{+\bullet}$ leads to a gradual loss in the molecular aromaticity through the breaking of the double bond in the aromatic ring of the $HMB^{+\bullet}$.[40] Consequently, the bond order involving the C-N-C and C-S-C moieties should decrease, leading to a shift to lower wavenumbers of their Raman vibrational modes, which is observed in Figure 1 B, C, and D for the spectral features at 1622, 1395, 1070 and 1040 $cm^{-1}$. Additionally, a new SERS band at 1500 $cm^{-1}$ appears in the intermediate, $HMB^{+\bullet}$ form. For the LMB form, the SERS intensity decreases drastically (Figure 1B), due to the loss of resonance in the visible part of electromagnetic spectrum.

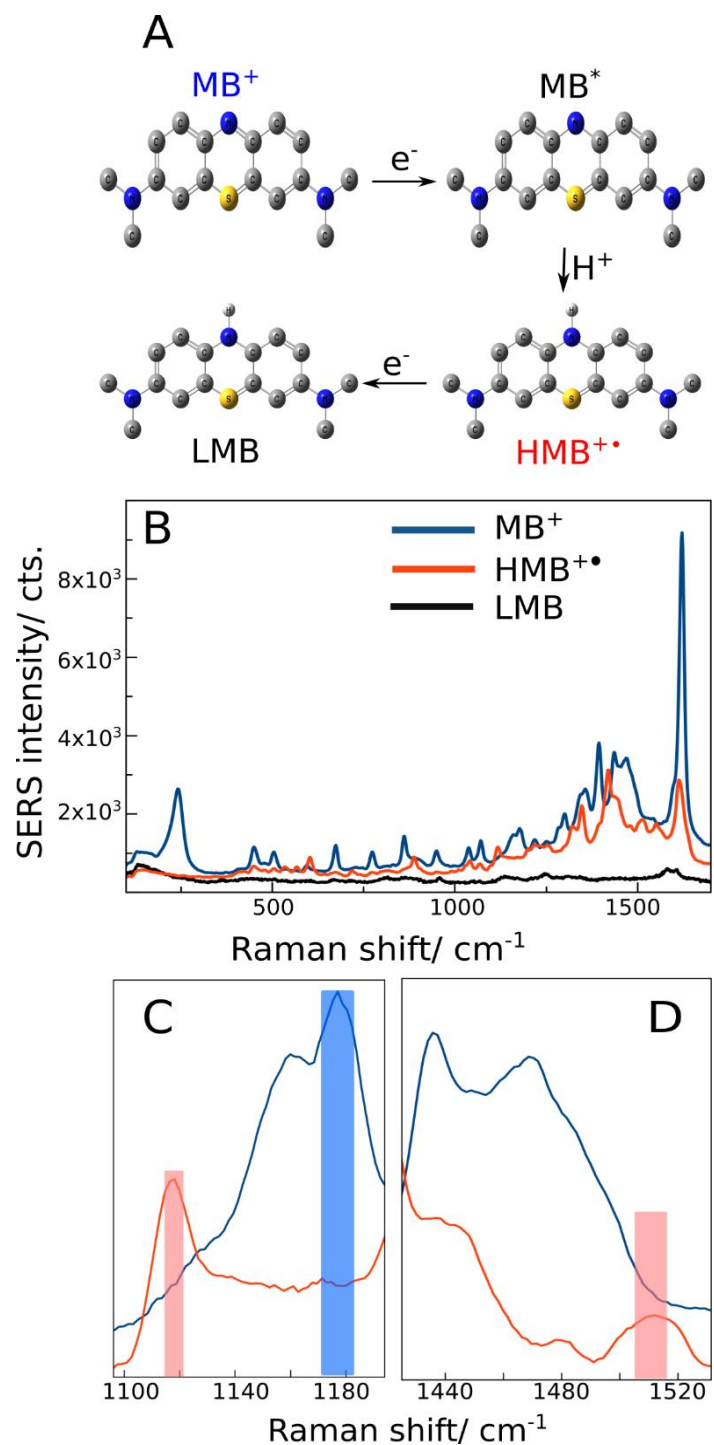

**Figure 1. (A)** The two-step reduction scheme of MB$^+$ to the first unstable intermediate form (MB*), to the radical intermediate form (HMB$^{+\bullet}$) and to the fully reduced form (LMB). **(B)** SERS spectra of MB$^+$, HMB$^{+\bullet}$ and LMB forms during the AgNPs-catalyzed reduction of MB$^+$ by NaBH$_4$, acquired with 532 nm excitation. **(C)** The appearance of the N-H bending mode on the intermediate HMB$^{+\bullet}$ form at 1130 cm$^{-1}$ and **(D)** C-C stretching mode of the thiazine ring at 1510 cm$^{-1}$ in the SERS spectrum of HMB$^{+\bullet}$.

Having discussed the MB$^+$ reduction and the SERS spectral changes associated with it, we will now turn our attention to the AgNPs used, and the Fermi energy upshift, *in-situ*, through chemisorbed halide ions.

**AgNPs characterization.** For the SERS measurements, we used citrate-capped AgNPs which we then modified with Cl$^-$ (AgNPs@Cl), Br$^-$ (AgNPs@Br) or I$^-$ (AgNPs@I). The halide ions have a higher affinity for the Ag surface, being able to displace the citrate capping agent and chemisorb to the metal surface (Supporting Information S3), influencing thus the electronic band structure of the AgNPs.

TEM images show highly monodispersed, spherical AgNPs with a 34±3.4 nm diameter (Figure 2A, B). The SEM images, together with UV-Vis and DLS spectra, suggest that no large aggregates form during the addition of 10 μM Br$^-$ or I$^-$, or 1 mM Cl$^-$ and MB$^+$ (0.45 μM) to the citrate-capped AgNPs (Supporting Information, Figures S4-S5; additional details on the adsorption of MB$^+$ to AgNPs can be found in Supporting Information, Figure S6).

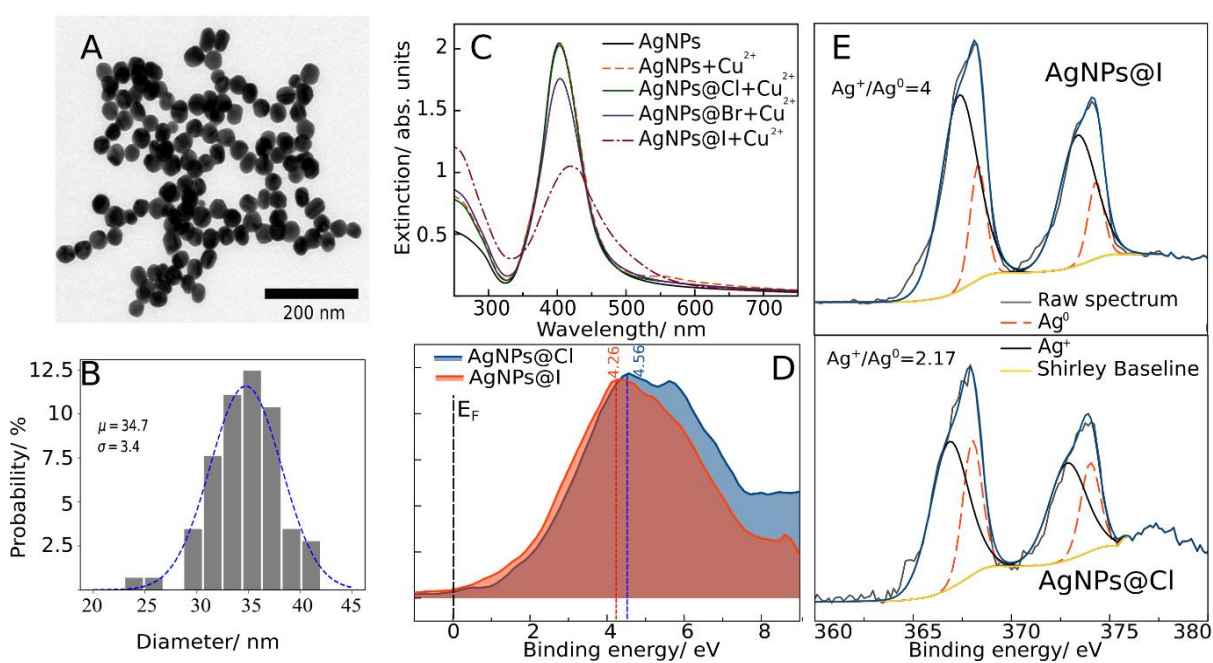

**Figure 2. (A)** A representative TEM micrograph of the AgNPs; **(B)** Size distribution of the AgNPs, determined from TEM images (n=72 particles) indicating a mean diameter of 34 nm; **(C)** Oxidative dissolution of AgNPs@Br and AgNPs@I in the presence of 1 mM final concentration of $CuSO_4$, observed through the decrease of their SPR band intensity. On the contrary, the addition of $Cu^{2+}$ to citrate- or AgNPs@Cl lets the SPR band unmodified compared to the as synthesized citrate capped AgNPs; **(D)** The valence band of AgNPs@Cl and AgNPs@I, the later showing a shift of 0.3 eV towards the Fermi level. The narrowing of the AgNPs@I valence band compared to the valence band of AgNPs@Cl can be explained by a strong depletion of the 5s orbital due to Ag oxidation. **(E)** Ag 3d doublet XPS spectral region of AgNPs@Cl and AgNPs@I, indicating a mixed-phase thin film consisting of both silver metal and oxidized silver. The band deconvolution shows a higher contribution of oxidized silver in case of AgNPs@I.

The strong adsorption of halide ions on AgNPs changes their electronic band structure, influencing their optical and electronic (catalytic) properties. In the case of an oxide for example, the Fermi energy, and consequently the work function, can be tuned by introducing oxygen defects.[43] Removing oxygen atoms leads to a decrease in the oxide's electron chemical potential. Oxygen vacancies act as n-type dopants and raise the Fermi level, decreasing the work function.[17] Analogously, we used Br⁻ and I⁻ that act similar to n-type dopants for AgNPs and raise the Fermi energy of AgNPs.[44-45] Due to the high nucleophilicity of Br⁻ and I⁻, during their adsorption on AgNPs, a partial charge transfer from the halide ion to the AgNP occurs, thus upshifting the Fermi energy of the metal.[24, 46]

By upshifting the Fermi energy of AgNPs in solution through the adsorption of Br⁻ and I⁻, AgNPs are more prone to oxidation and dissolution by species that would not react with them normally, such as $Cu^{2+}$.[47-48] Figure 2C shows that the SPR band intensity decreases significantly only for AgNPs@Br and AgNPs@I due to the upshift of their Fermi energy high enough to drive the reduction of $Cu^{2+}$ to $Cu^+$ (0.16 V standard reduction potential), and the subsequent oxidation of $Ag^0$ to $Ag^+$, leading to the dissolution of AgNPs.[47, 49] By replacing $CuSO_4$ with $Na_2SO_4$, no change to the SPR of colloidal AgNPs was observed, since $Na^+$ has a much higher standard reduction potential (-2.71 V), and therefore does not react with the AgNPs after the addition of halide ions (Supporting Information, Figure S7). The upshift of the Fermi level of AgNPs was further confirmed by DFT calculations which showed an upshift of 0.16 and 0.47 eV of AgNPs@Br and AgNPs@I, respectively, compared to AgNPs@Cl (Supporting Information, Figures S8-S9).

XPS spectroscopy can also reveal changes in the electronic band structure of AgNPs, by probing binding energies of core and valence electrons in AgNPs. Figure 2 D and E show the experimental binding energies of the valence band and of the $3d_{5/2}$ and $3d_{3/2}$ spin-orbit partners (368 and 374 eV), respectively, for the two extreme cases in term of Fermi energy change, AgNPs@Cl and AgNPs@I. The upshift of the 4d valence band by 0.3 eV, as well as the increase of the $Ag^+/Ag^0$ ratio on AgNPs@I obtained from the 3d doublet deconvolution, indicate a charge transfer form $I^-$ to the Ag surface and consequently an increase of the amount of oxidized silver of the AgNPs@I compared to AgNPs@Cl, due to the subsequent oxidation of $Ag^0$ by ambient oxygen.[24, 46, 50-55] Similar shifts of Fermi level up to ~0.3 eV, induced by $Br^-$ and $I^-$, were observed for silver-perovskite nanocrystals by S. Saris *et. al.*[56]

The reversibility of $Cl^-$ and $Br^-$ adsorption (Supporting Information, Figure S3) indicates that there is no mixed-phase AgCl or AgBr forming and points towards a surface modification of the AgNPs rather than surface catalysis, while the SEM, DLS and UV-Vis measurements exclude any important morphological changes of the AgNPs after the addition of halide ions and $MB^+$ (Supporting Information, Figures S4 and S5).

**Energy level equilibration.** In this section we present the main contributions of the paper. Figure 3 shows the formation of HMB$^{+\bullet}$ species on the surface of AgNPs@Br and AgNPs@I, tracked by SERS. Figure 3B shows the SERS bands at 1150 (from the N-H bond of HMB$^{+\bullet}$) and 1180 cm$^{-1}$ (from the MB$^{+}$) on AgNPs@Cl, AgNPs@Br and AgNPs@I. By taking their relative ratio, we can infer semi-quantitatively the ratio of the two molecular forms on the three different Ag surfaces (Figure 3E). To correlate the HMB$^{+\bullet}$/MB$^{+}$ ratio with the Fermi level upshift on AgNPs@Br and AgNPs@I, we plotted the 1150/1180 cm$^{-1}$ SERS intensity (i.e., % of conversion) against the DFT calculated Fermi level (on the x axis) of the three metal surfaces (Figure 3E). Additional spectral changes corresponding to HMB$^{+\bullet}$ formation are shown in Supporting Information, Figure S10.

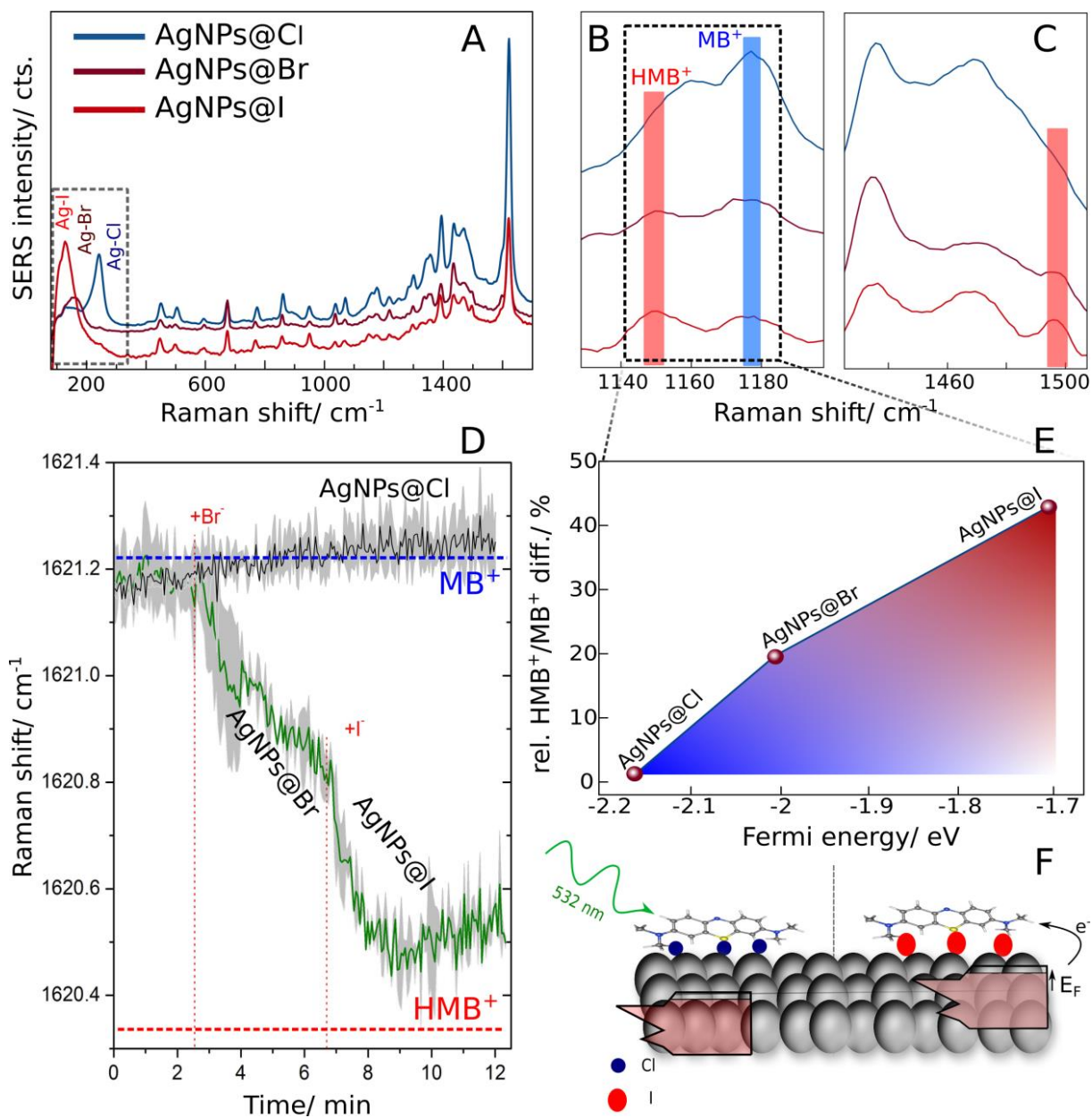

**Figure 3.** The reduction of MB$^+$ on the surface of AgNPs, due to the shift of the Fermi level, monitored with 532 nm excitation. **(A)** SERS spectra of MB$^+$ on AgNPs@Cl, AgNPs@Br and AgNPs@I. The low wavenumber region shows the surface dynamics of Cl$^-$ (240 cm$^{-1}$), Br$^-$ (160 cm$^{-1}$) and I$^-$ (130 cm$^{-1}$). **(B-C)** Insets from (A), showing the spectral changes characteristic to HMB$^{+\bullet}$ that appear after the adsorption of Br$^-$ and especially I$^-$. **(D)** Time-series SERS measurements of MB$^+$ on AgNPs@Cl, modifying, *in-situ*, the surface of AgNPs with Br$^-$ and I$^-$ (10 µM), at the marked times. After the sequential addition of Br$^-$ and I$^-$, the SERS peak at 1621 cm$^{-1}$ shifts to lower wavenumbers, characteristic to HMB$^{+\bullet}$. **(E)** The relative percentual

increase in the 1150/1180 cm$^{-1}$ SERS intensity ratio corresponding to the HMB$^{+\bullet}$/ MB$^+$ ratio on the surface of AgNPs@Cl, AgNPs@Br and AgNPs@I from (B) plotted against the Fermi level energies determined from the DFT calculations, on the x-axis. **(F)** Schematic representation of the reduction of MB$^+$ on the surface of AgNPs, due to the shift of Fermi level after the chemisorption of I$^-$. Analogous to electrochemistry experiments, the Fermi level of AgNPs upshifts, favoring the *spontaneous* electron transfer across the interface to the unoccupied level of adsorbed MB$^+$.

Importantly, the Ag-halide SERS band at low wavenumbers allow us to monitor the surface dynamics of halide ions and MB$^+$ in real time, gaining thus a mechanistic understanding on the elusive role of halides in SERS and plasmonic chemistry (Figure 3A).

Taking advantage of the possibility of tracking *in-situ* the surface dynamics of halide ions and MB$^+$, we monitored the exchange of halide ions on the surface of AgNPs, and the consequent reduction of MB$^+$ in Figure 3D. At t = 0, MB$^+$ is chemisorbed on AgNPs@Cl, the MB$^+$ adsorption being mediated by Cl$^-$.[57-59] At t = 2.5 min, KBr was added to the colloidal solution and Br$^-$ replaces Cl$^-$ on the surface of the AgNPs due to its higher surface affinity (Supporting Information, Figure S3). Adsorbed Br$^-$ upshifts the Fermi energy of AgNPs and promotes the charge transfer from the Fermi energy to the unoccupied orbital of adsorbed MB$^+$, forming HMB$^{+\bullet}$. At t ≈ 6 min, KI was added to the colloidal solution, upshifting the Fermi energy of AgNPs even more and promoting further the charge transfer from the Fermi level to the LUMO orbital of MB$^+$ until the energy levels are equilibrated, as evidenced by the plateau reached at the end of the measurement (10–12 min). Solely in the presence of Cl$^-$, the 1621 cm$^{-1}$ band remains constant (Figure 3D), proving that the reduction of MB$^+$ is thermodynamically driven by the equilibration of the energy levels of the AgNPs–MB$^+$ system, and not due to the photoexcitation of electrons from the AgNPs upon 532 nm excitation or from the heating of AgNPs. The decrease in intensity of MB$^+$ SERS bands as Br$^-$ and I$^-$ are added (Figure 3A) is attributed to the partial reduction of MB$^+$ to the less radiative form HMB$^{+\bullet}$ (Supporting Information, Figure S11).[35] Additional experimental evidence excluding the photo- or thermal reduction of MB$^+$ can be found in Supporting Information, Figures S14-S17, as well as in previous reports.[32-34]

To confirm these experimental results, we also calculated the density of states for AgNP-MB$^+$ system in the three cases (i.e., AgNPs@Cl, AgNPs@Br and AgNPs@I). The results show that the LUMO orbital of MB$^+$ is close to the Fermi level of AgNPs@Cl; as the Fermi level upshifts due to the adsorption of Br$^-$ and I$^-$, it crosses the unoccupied orbital of MB$^+$, populating it with electrons, thus forming the HMB$^{+\bullet}$ form (Supporting Information, Figures S12 and S13).

As shown in previous publications, MB$^+$ chemisorbs to the AgNPs surface forming hybrid metal-MB$^+$ orbitals associated with the whole AgNP-MB$^+$ complex.[60] This can additionally favor the metal to MB$^+$ charge transfer since the Fermi level of the AgNP does not have to be shifted by the energy required for the heterogeneous reduction of MB$^+$ to HMB$^{+\bullet}$ (-0.27 V)[36,40]. Instead, the Fermi level must now cross the new (hybridized) unoccupied orbital of the adsorbed MB$^+$, which can be at lower energies than the unoccupied orbital of free MB$^+$, as shown in previous reports, increasing the efficiency of charge transfer.[32-33]

Due to the richness and depth of the halide-assisted metal-molecule charge transfer effects, we thought to probe other systems as well, to see if the Fermi level shift has the same impact on other metal-adsorbate systems.

We found that halide ions can play different roles in interfacial charge transfer. When molecules with a higher affinity for the silver surface than that of halide ions (such as many thiolated molecules) are introduced in the system, the halide ions cannot directly adsorb onto the metal surface and influence the electronic levels of the AgNPs, as in the AgNP–MB$^+$ system. Instead, in such systems the halide ions can act just as hole scavengers for the hot holes created through photoexcitation, transiently adsorbing on the AgNPs (Figure 4A-B). To probe our hypothesis we chose the reduction of 4-nitrothiophenol (NTP) to aminothiophenol (ATP) as a model system.[61-62] No Ag-halide SERS band can be observed in the SERS spectra of NTP, indicating that the halide ions are not co-adsorbed with NTP (Supporting Information, S18).

To probe the time-dependent interaction of I$^-$ with the AgNPs, we monitored the Ag-I SERS band (130 cm$^{-1}$) and the Ag-S SERS band of NTP (200 cm$^{-1}$) through time-series SERS.[63] During the time-series measurement we could observe a few events where the Ag-I band appeared very intensely, after which it disappeared, whereas the Ag-S band remained constant throughout the measurement. Importantly, in the SERS acquisitions where the Ag-I band was present, we also observed the two SERS bands at 1570 and 1590 cm$^{-1}$ SERS band, characteristic to NTP reduction to ATP (Figure 4C and D).

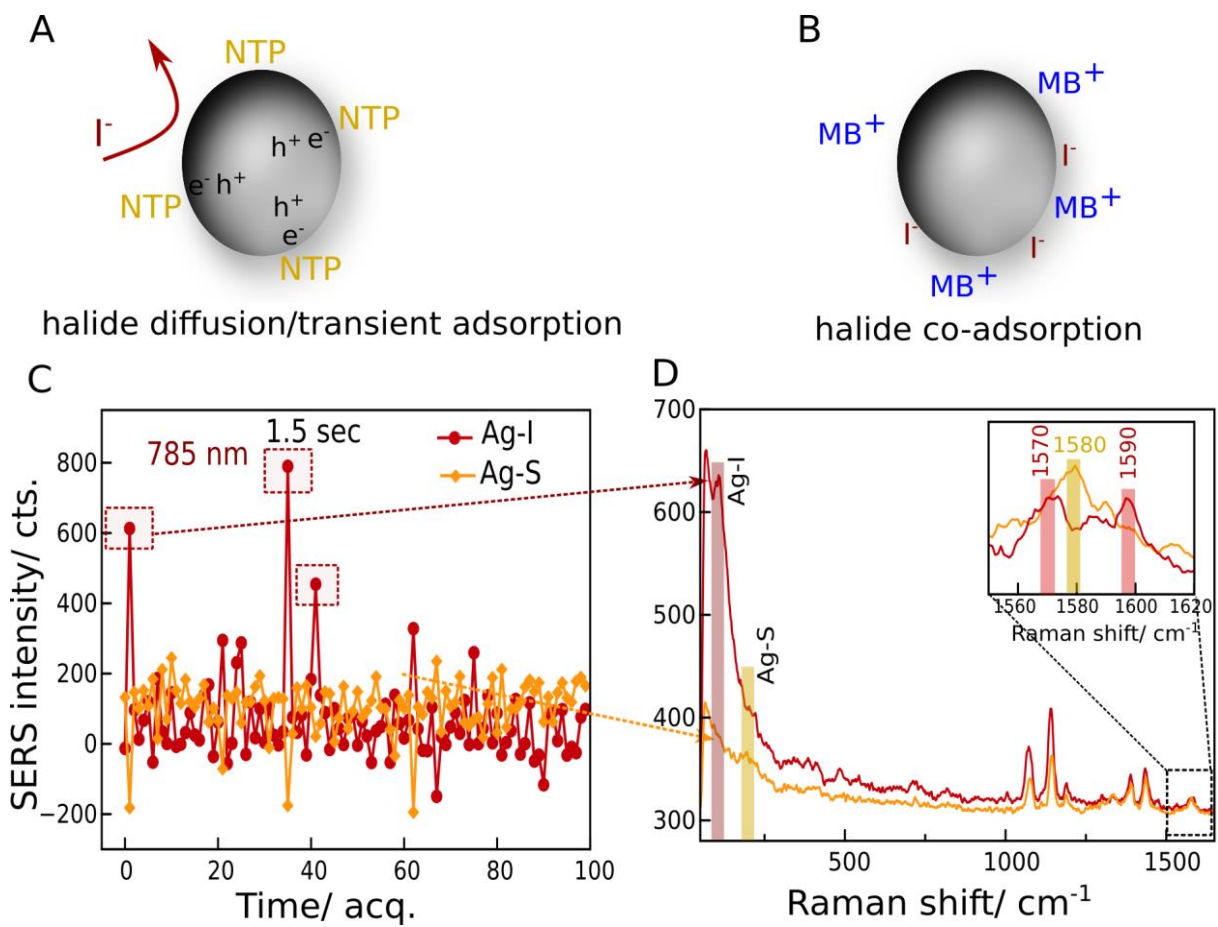

**Figure 4.** Schematic representation of the **(A)** transient adsorption of I⁻ on AgNPs with NTP, suggested by the lack of the Ag-I SERS band and **(B)** co-adsorption of MB⁺ and I⁻ ions on AgNPs. **(C)** Time-series SERS monitoring (100 acquisitions, 1.5 sec. integration time) of the I⁻ transient adsorption during excitation with 785 nm and **(D)** corresponding SERS spectra of NTP. The inset in D shows the appearance of the 1570 and 1590 cm$^{-1}$ SERS bands, corresponding to the reduction of NTP to ATP.

Although the SERS peaks of DMAB, at ~ 1100, 1380 and 1430 cm$^{-1}$ can be observed, the appearance of the two peaks at 1570 and 1590 cm$^{-1}$ is a telltale sign of the conversion of NTP to ATP and is not observed in DMAB[61-62]. Moreover, the SERS peaks of DMAB were constant throughout the time-series measurement (Figure 4D), while the two SERS peaks at 1570 and 1590 cm$^{-1}$ appeared only correlated with the Ag-I SERS band (as shown in the inset of Figure 4D), further suggesting that I⁻ catalyzed the conversion of NTP to ATP by transiently adsorbing to the surface of AgNPs.

In conclusion, we have shown that the role of halide ions in the metal–molecule charge transfer process is dictated by the adsorption regime of the halide ions. Through the chemisorption of halide ions, Ag-X (X= Cl⁻, Br⁻, I⁻) surface complexes form, which alter the surface potential and Fermi energy of the AgNPs. This, in turn, can lead to spontaneous metal-molecule charge transfer if the acceptor orbitals of the adsorbed molecule are properly aligned with the Fermi level of the metal, as in the AgNPs-MB⁺ case. Our results show that the effect of adsorbed ions on plasmonic nanomaterials is energetically comparable (~0.3 eV) with the light-induce surface voltage effect (photovoltage) or the energy of hot-carriers participating in catalytic processes.[13, 31, 37-39]

However, when the metal surface is occupied with molecules with a stronger affinity for the metal surface than halide ions such as thiolated molecules, the halide ions cannot co-adsorb on the surface, which can be readily checked experimentally through the Ag-halide SERS bands (also valid for Au NPs)[64-67]. Thus, in these systems halides' catalytic role involves scavenging the hot holes created through SPR excitations or inter- and intraband transitions[61-62]. These results highlight the important and many times underestimated role of capping agents, electric double layer, reaction media and halides co-adsorption in plasmonic catalysis and surface-enhanced spectroscopies.

**Supporting Information** is available free of charge. The content includes AgNPs synthesis and additional experimental details, UV-Vis, DLS and SEM characterization of AgNPs; SERS measurements of $MB^+$ and DFT calculations of the Fermi level and density of states of the AgNPs-$MB^+$ system.

### Acknowledgements

We are thankful to Professor Simion Astilean for useful discussions and for reading the manuscript. Matias Herran is highly acknowledged for the TEM images. Andrei Stefancu is thankful to an Erasmus+ scholarship for facilitating his research placement at LMU Munich.

### Funding

We acknowledge financial support from the Romanian Ministry of Research and Innovation, CCCDI-UEFISCDI, project numbers PN-III-P1-1.2-PCCDI-2017-0056, PN-III-P2-2.1-PED-2019-3268 and PN-III-P4-ID-PCCF-2016-0112 within PNCDI III. We also acknowledge funding and support from the Deutsche Forschungsgemeinschaft (DFG, German Research Foundation) under Germany´s Excellence Strategy – EXC 2089/1 – 390776260, the Bavarian program Solar Energies Go Hybrid (SolTech), the Center for NanoScience (CeNS) and the European Commission through the ERC Starting Grant CATALIGHT (802989).

### Notes

The authors declare no competing financial interest.

### References

1.      Gargiulo, J.; Berté, R.; Li, Y.; Maier, S. A.; Cortés, E., From Optical to Chemical Hot Spots in Plasmonics. *Accounts of Chemical Research* **2019,** *52* (9), 2525-2535.
2.      Brongersma, M. L.; Halas, N. J.; Nordlander, P., Plasmon-induced hot carrier science and technology. *Nature Nanotechnology* **2015,** *10* (1), 25-34.
3.      Linic, S.; Christopher, P.; Ingram, D. B., Plasmonic-metal nanostructures for efficient conversion of solar to chemical energy. *Nature Materials* **2011,** *10* (12), 911-921.

4. Zhou, L.; Martirez, J. M. P.; Finzel, J.; Zhang, C.; Swearer, D. F.; Tian, S.; Robatjazi, H.; Lou, M.; Dong, L.; Henderson, L.; Christopher, P.; Carter, E. A.; Nordlander, P.; Halas, N. J., Light-driven methane dry reforming with single atomic site antenna-reactor plasmonic photocatalysts. *Nature Energy* **2020,** *5* (1), 61-70.
5. Yu, S.; Wilson, A. J.; Heo, J.; Jain, P. K., Plasmonic Control of Multi-Electron Transfer and C–C Coupling in Visible-Light-Driven CO2 Reduction on Au Nanoparticles. *Nano Letters* **2018,** *18* (4), 2189-2194.
6. Christopher, P.; Xin, H.; Linic, S., Visible-light-enhanced catalytic oxidation reactions on plasmonic silver nanostructures. *Nature Chemistry* **2011,** *3* (6), 467-472.
7. Stefancu, A.; Iancu, S.; Leopold, L.; Leopold, N., Contribution of chemical interface damping to the shift of surface plasmon resonance energy of gold nanoparticles. *Romanian Reports in Physics* **2020,** *72* (1), 402.
8. Cortés, E., Efficiency and Bond Selectivity in Plasmon-Induced Photochemistry. *Advanced Optical Materials* **2017,** *5* (15), 1700191.
9. Cortés, E.; Besteiro, L. V.; Alabastri, A.; Baldi, A.; Tagliabue, G.; Demetriadou, A.; Narang, P., Challenges in Plasmonic Catalysis. *ACS Nano* **2020,** *14* (12), 16202-16219.
10. Devasia, D.; Das, A.; Mohan, V.; Jain, P. K., Control of Chemical Reaction Pathways by Light–Matter Coupling. *Annual Review of Physical Chemistry* **2021,** *72*, 423-443.
11. Lee, S.; Hwang, H.; Lee, W.; Schebarchov, D.; Wy, Y.; Grand, J.; Auguié, B.; Wi, D. H.; Cortés, E.; Han, S. W., Core–Shell Bimetallic Nanoparticle Trimers for Efficient Light-to-Chemical Energy Conversion. *ACS Energy Letters* **2020,** *5* (12), 3881-3890.
12. Kamarudheen, R.; Aalbers, G. J. W.; Hamans, R. F.; Kamp, L. P. J.; Baldi, A., Distinguishing Among All Possible Activation Mechanisms of a Plasmon-Driven Chemical Reaction. *ACS Energy Letters* **2020,** *5* (8), 2605-2613.
13. Pensa, E.; Gargiulo, J.; Lauri, A.; Schlücker, S.; Cortés, E.; Maier, S. A., Spectral Screening of the Energy of Hot Holes over a Particle Plasmon Resonance. *Nano Letters* **2019,** *19* (3), 1867-1874.
14. Aslam, U.; Chavez, S.; Linic, S., Controlling energy flow in multimetallic nanostructures for plasmonic catalysis. *Nature Nanotechnology* **2017,** *12* (10), 1000-1005.
15. Hollerer, M.; Lüftner, D.; Hurdax, P.; Ules, T.; Soubatch, S.; Tautz, F. S.; Koller, G.; Puschnig, P.; Sterrer, M.; Ramsey, M. G., Charge Transfer and Orbital Level Alignment at Inorganic/Organic Interfaces: The Role of Dielectric Interlayers. *ACS Nano* **2017,** *11* (6), 6252-6260.
16. Brown, P. R.; Kim, D.; Lunt, R. R.; Zhao, N.; Bawendi, M. G.; Grossman, J. C.; Bulović, V., Energy Level Modification in Lead Sulfide Quantum Dot Thin Films through Ligand Exchange. *ACS Nano* **2014,** *8* (6), 5863-5872.
17. Greiner, M. T.; Helander, M. G.; Tang, W.-M.; Wang, Z.-B.; Qiu, J.; Lu, Z.-H., Universal energy-level alignment of molecules on metal oxides. *Nature Materials* **2012,** *11* (1), 76-81.
18. Yang, X.; Egger, L.; Fuchsberger, J.; Unzog, M.; Lüftner, D.; Hajek, F.; Hurdax, P.; Jugovac, M.; Zamborlini, G.; Feyer, V.; Koller, G.; Puschnig, P.; Tautz, F. S.; Ramsey, M. G.; Soubatch, S., Coexisting Charge States in a Unary Organic Monolayer Film on a Metal. *The Journal of Physical Chemistry Letters* **2019,** *10* (21), 6438-6445.
19. Sundaresan, V.; Monaghan, J. W.; Willets, K. A., Visualizing the Effect of Partial Oxide Formation on Single Silver Nanoparticle Electrodissolution. *The Journal of Physical Chemistry C* **2018,** *122* (5), 3138-3145.
20. Al-Zubeidi, A.; Hoener, B. S.; Collins, S. S. E.; Wang, W.; Kirchner, S. R.; Hosseini Jebeli, S. A.; Joplin, A.; Chang, W.-S.; Link, S.; Landes, C. F., Hot Holes Assist Plasmonic Nanoelectrode Dissolution. *Nano Letters* **2019,** *19* (2), 1301-1306.
21. Azodi, M.; Sultan, Y.; Ghoshal, S., Dissolution Behavior of Silver Nanoparticles and Formation of Secondary Silver Nanoparticles in Municipal Wastewater by Single-Particle ICP-MS. *Environmental Science & Technology* **2016,** *50* (24), 13318-13327.

22. Henglein, A., Physicochemical properties of small metal particles in solution: "microelectrode" reactions, chemisorption, composite metal particles, and the atom-to-metal transition. *The Journal of Physical Chemistry* **1993,** *97* (21), 5457-5471.
23. Henglein, A., Colloidal Silver Nanoparticles:  Photochemical Preparation and Interaction with O2, CCl4, and Some Metal Ions. *Chemistry of Materials* **1998,** *10* (1), 444-450.
24. Linnert, T.; Mulvaney, P.; Henglein, A., Surface chemistry of colloidal silver: surface plasmon damping by chemisorbed iodide, hydrosulfide (SH-), and phenylthiolate. *The Journal of Physical Chemistry* **1993,** *97* (3), 679-682.
25. Stadler, R.; Jacobsen, K. W., Fermi level alignment in molecular nanojunctions and its relation to charge transfer. *Physical Review B* **2006,** *74* (16), 161405.
26. Tang, J.; Kemp, K. W.; Hoogland, S.; Jeong, K. S.; Liu, H.; Levina, L.; Furukawa, M.; Wang, X.; Debnath, R.; Cha, D.; Chou, K. W.; Fischer, A.; Amassian, A.; Asbury, J. B.; Sargent, E. H., Colloidal-quantum-dot photovoltaics using atomic-ligand passivation. *Nature Materials* **2011,** *10* (10), 765-771.
27. Lin, K.; Xing, J.; Quan, L. N.; de Arquer, F. P. G.; Gong, X.; Lu, J.; Xie, L.; Zhao, W.; Zhang, D.; Yan, C.; Li, W.; Liu, X.; Lu, Y.; Kirman, J.; Sargent, E. H.; Xiong, Q.; Wei, Z., Perovskite light-emitting diodes with external quantum efficiency exceeding 20 per cent. *Nature* **2018,** *562* (7726), 245-248.
28. Sundararaman, R.; Narang, P.; Jermyn, A. S.; Goddard Iii, W. A.; Atwater, H. A., Theoretical predictions for hot-carrier generation from surface plasmon decay. *Nature Communications* **2014,** *5* (1), 5788.
29. Manjavacas, A.; Liu, J. G.; Kulkarni, V.; Nordlander, P., Plasmon-Induced Hot Carriers in Metallic Nanoparticles. *ACS Nano* **2014,** *8* (8), 7630-7638.
30. Ostovar, B.; Cai, Y.-Y.; Tauzin, L. J.; Lee, S. A.; Ahmadivand, A.; Zhang, R.; Nordlander, P.; Link, S., Increased Intraband Transitions in Smaller Gold Nanorods Enhance Light Emission. *ACS Nano* **2020,** *14* (11), 15757-15765.
31. Reddy, H.; Wang, K.; Kudyshev, Z.; Zhu, L.; Yan, S.; Vezzoli, A.; Higgins, S. J.; Gavini, V.; Boltasseva, A.; Reddy, P.; Shalaev, V. M.; Meyhofer, E., Determining plasmonic hot-carrier energy distributions via single-molecule transport measurements. *Science* **2020,** *369* (6502), 423.
32. Boerigter, C.; Aslam, U.; Linic, S., Mechanism of Charge Transfer from Plasmonic Nanostructures to Chemically Attached Materials. *ACS Nano* **2016,** *10* (6), 6108-6115.
33. Boerigter, C.; Campana, R.; Morabito, M.; Linic, S., Evidence and implications of direct charge excitation as the dominant mechanism in plasmon-mediated photocatalysis. *Nature Communications* **2016,** *7*, 10545.
34. Rao, V. G.; Aslam, U.; Linic, S., Chemical Requirement for Extracting Energetic Charge Carriers from Plasmonic Metal Nanoparticles to Perform Electron-Transfer Reactions. *Journal of the American Chemical Society* **2019,** *141* (1), 643-647.
35. Tognalli, N. G.; Fainstein, A.; Vericat, C.; Vela, M. E.; Salvarezza, R. C., In Situ Raman Spectroscopy of Redox Species Confined in Self-Assembled Molecular Films. *The Journal of Physical Chemistry C* **2008,** *112* (10), 3741-3746.
36. Nicolai, S. H. A.; Rubim, J. C., Surface-Enhanced Resonance Raman (SERR) Spectra of Methylene Blue Adsorbed on a Silver Electrode. *Langmuir* **2003,** *19* (10), 4291-4294.
37. Sheldon, M. T.; van de Groep, J.; Brown, A. M.; Polman, A.; Atwater, H. A., Plasmoelectric potentials in metal nanostructures. *Science* **2014,** *346* (6211), 828.
38. Wilson, A. J.; Jain, P. K., Light-Induced Voltages in Catalysis by Plasmonic Nanostructures. *Accounts of Chemical Research* **2020,** *53* (9), 1773-1781.
39. Kim, Y.; Dumett Torres, D.; Jain, P. K., Activation Energies of Plasmonic Catalysts. *Nano Letters* **2016,** *16* (5), 3399-3407.
40. Nicolai, S. l. H. d. A.; Rodrigues, P. R. P.; Agostinho, S. M. L.; Rubim, J. C., Electrochemical and spectroelectrochemical (SERS) studies of the reduction of methylene blue on a silver electrode. *Journal of Electroanalytical Chemistry* **2002,** *527* (1), 103-111.
41. Kamat, P. V.; Lichtin, N. N., Photoinduced electron-ejection from methylene blue in water and acetonitrile. *The Journal of Physical Chemistry* **1981,** *85* (25), 3864-3868.

42. Jiang, Z.-J.; Liu, C.-Y.; Sun, L.-W., Catalytic Properties of Silver Nanoparticles Supported on Silica Spheres. *The Journal of Physical Chemistry B* **2005,** *109* (5), 1730-1735.
43. Glass, D.; Cortés, E.; Ben-Jaber, S.; Brick, T.; Peveler, W. J.; Blackman, C. S.; Howle, C. R.; Quesada-Cabrera, R.; Parkin, I. P.; Maier, S. A., Dynamics of Photo-Induced Surface Oxygen Vacancies in Metal-Oxide Semiconductors Studied Under Ambient Conditions. *Advanced Science* **2019,** *6* (22), 1901841.
44. Mantri, Y.; Davidi, B.; Lemaster, J. E.; Hariri, A.; Jokerst, J. V., Iodide-doped precious metal nanoparticles: measuring oxidative stress in vivo via photoacoustic imaging. *Nanoscale* **2020,** *12* (19), 10511-10520.
45. Zapata Herrera, M.; Aizpurua, J.; Kazansky, A. K.; Borisov, A. G., Plasmon Response and Electron Dynamics in Charged Metallic Nanoparticles. *Langmuir* **2016,** *32* (11), 2829-2840.
46. Marichev, V. A., First experimental evaluation of partial charge transfer during anion adsorption. *Colloids and Surfaces A: Physicochemical and Engineering Aspects* **2009,** *348* (1), 28-34.
47. Li, N.; Gu, J.; Cui, H., Luminol chemiluminescence induced by silver nanoparticles in the presence of nucleophiles and Cu2+. *Journal of Photochemistry and Photobiology A: Chemistry* **2010,** *215* (2), 185-190.
48. Scanlon, M. D.; Peljo, P.; Méndez, M. A.; Smirnov, E.; Girault, H. H., Charging and discharging at the nanoscale: Fermi level equilibration of metallic nanoparticles. *Chemical Science* **2015,** *6* (5), 2705-2720.
49. Stefancu, A.; Iancu, S. D.; Coman, V.; Leopold, L. F.; Leopold, N., Tuning the potential of nanoelectrodes to maximum: Ag and Au nanoparticles dissolution by I- adsorption via Mg2+ adions. *Romanian Reports in Physics* **2021,** *73* (2), 501.
50. Kaspar, T. C.; Droubay, T.; Chambers, S. A.; Bagus, P. S., Spectroscopic Evidence for Ag(III) in Highly Oxidized Silver Films by X-ray Photoelectron Spectroscopy. *The Journal of Physical Chemistry C* **2010,** *114* (49), 21562-21571.
51. Al-Hada, M.; Gregoratti, L.; Amati, M.; Neeb, M., Pristine and oxidised Ag-nanoparticles on free-standing graphene as explored by X-ray photoelectron and Auger spectroscopy. *Surface Science* **2020,** *693*, 121533.
52. Panaccione, G.; Cautero, G.; Cautero, M.; Fondacaro, A.; Grioni, M.; Lacovig, P.; Monaco, G.; Offi, F.; Paolicelli, G.; Sacchi, M.; Stojić, N.; Stefani, G.; Tommasini, R.; Torelli, P., High-energy photoemission in silver: resolving d and sp contributions in valence band spectra. *Journal of Physics: Condensed Matter* **2005,** *17* (17), 2671-2679.
53. Ghosalya, M. K.; Reddy, K. P.; Jain, R.; Roy, K.; Gopinath, C. S., Subtle interaction between Ag and $$\hbox {O}_{2}$$O2: a near ambient pressure UV photoelectron spectroscopy (NAP-UPS) investigations. *Journal of Chemical Sciences* **2018,** *130* (3), 30.
54. Lee, Y. W.; Ahn, H.; Lee, S. E.; Woo, H.; Han, S. W., Fine Control over the Compositional Structure of Trimetallic Core–Shell Nanocrystals for Enhanced Electrocatalysis. *ACS Applied Materials & Interfaces* **2019,** *11* (29), 25901-25908.
55. Chang, Y.; Cheng, Y.; Feng, Y.; Li, K.; Jian, H.; Zhang, H., Upshift of the d Band Center toward the Fermi Level for Promoting Silver Ion Release, Bacteria Inactivation, and Wound Healing of Alloy Silver Nanoparticles. *ACS Applied Materials & Interfaces* **2019,** *11* (13), 12224-12231.
56. Saris, S.; Niemann, V.; Mantella, V.; Loiudice, A.; Buonsanti, R., Understanding the mechanism of metal-induced degradation in perovskite nanocrystals. *Nanoscale* **2019,** *11* (41), 19543-19550.
57. Leopold, N.; Stefancu, A.; Herman, K.; Tódor, I. S.; Iancu, S. D.; Moisoiu, V.; Leopold, L. F., The role of adatoms in chloride-activated colloidal silver nanoparticles for surface-enhanced Raman scattering enhancement. *Beilstein Journal of Nanotechnology* **2018,** *9*, 2236-2247.
58. Doering, W. E.; Nie, S., Single-Molecule and Single-Nanoparticle SERS: Examining the Roles of Surface Active Sites and Chemical Enhancement. *The Journal of Physical Chemistry B* **2002,** *106* (2), 311-317.

59. Stefancu, A.; Iancu, S. D.; Leopold, N., Selective Single Molecule SERRS of Cationic and Anionic Dyes by Cl– and Mg2+ Adions: An Old New Idea. *The Journal of Physical Chemistry C* **2021,** *125* (23), 12802-12810.
60. Boerigter, C.; Campana, R.; Morabito, M.; Linic, S., Evidence and implications of direct charge excitation as the dominant mechanism in plasmon-mediated photocatalysis. *Nature Communications* **2016,** *7* (1), 10545.
61. Cortés, E.; Xie, W.; Cambiasso, J.; Jermyn, A. S.; Sundararaman, R.; Narang, P.; Schlücker, S.; Maier, S. A., Plasmonic hot electron transport drives nano-localized chemistry. *Nature Communications* **2017,** *8* (1), 14880.
62. Xie, W.; Schlücker, S., Hot electron-induced reduction of small molecules on photorecycling metal surfaces. *Nature Communications* **2015,** *6* (1), 7570.
63. Holze, R., The adsorption of thiophenol on gold – a spectroelectrochemical study. *Physical Chemistry Chemical Physics* **2015,** *17* (33), 21364-21372.
64. Perera, G. S.; LaCour, A.; Zhou, Y.; Henderson, K. L.; Zou, S.; Perez, F.; Emerson, J. P.; Zhang, D., Iodide-Induced Organothiol Desorption and Photochemical Reaction, Gold Nanoparticle (AuNP) Fusion, and SERS Signal Reduction in Organothiol-Containing AuNP Aggregates. *The Journal of Physical Chemistry C* **2015,** *119* (8), 4261-4267.
65. Athukorale, S.; De Silva, M.; LaCour, A.; Perera, G. S.; Pittman, C. U.; Zhang, D., NaHS Induces Complete Nondestructive Ligand Displacement from Aggregated Gold Nanoparticles. *The Journal of Physical Chemistry C* **2018,** *122* (4), 2137-2144.
66. Gao, P.; Weaver, M. J., Metal-adsorbate vibrational frequencies as a probe of surface bonding: halides and pseudohalides at gold electrodes. *The Journal of Physical Chemistry* **1986,** *90* (17), 4057-4063.
67. Chen, G.-Y.; Sun, Y.-B.; Shi, P.-C.; Liu, T.; Li, Z.-H.; Luo, S.-H.; Wang, X.-C.; Cao, X.-Y.; Ren, B.; Liu, G.-K.; Yang, L.-L.; Tian, Z.-Q., Revealing unconventional host–guest complexation at nanostructured interface by surface-enhanced Raman spectroscopy. *Light: Science & Applications* **2021,** *10* (1), 85.

Table of Content (TOC)

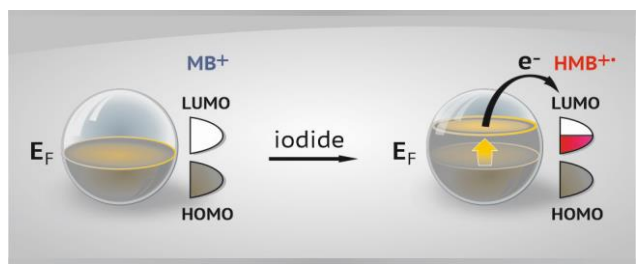